\documentstyle[prl,aps,twocolumn]{revtex}

\begin{document}

{\bf Comment on ``Teleportation with a uniformly accelerated partner''}

\bigskip

The authors of Ref.~\cite{Alsing} consider teleportation with one of
the partners (Rob) being uniformly accelerated and the other (Alice)
at rest and conclude that the corresponding fidelity is reduced due to  
the fact that an observer in a uniformly accelerated  frame (i.e., Rob) 
experiences the Minkowski vacuum (as seen by Alice) as a thermal bath.
However, the derivation presented in Ref.~\cite{Alsing} does not take 
into account several crucial features of the scenario under
investigation and hence the results of Ref.~\cite{Alsing} do not apply
in the general case -- instead the potential loss of fidelity strongly
depends on the explicit physical realization (see the points below). 
In particular, if cavities are used 
(which is the assumption of Ref.~\cite{Alsing}), the walls of the
cavities will keep out the thermal fluctuations 
(implying that there is no generic loss of fidelity). 

{\bf I.} Before Alice and Rob share the entangled state, one has to
specify how Rob removes all photons from his cavity -- in particular, 
which particle and vacuum definition Rob adopts in this process:
Does he define particles (instantaneously) with respect to the
Minkowski time or with respect to his proper (Rindler) time? 
The natural time coordinate inside an accelerated box is Rindler time: 
note that not placing the box into the Rindler vacuum state before
introducing the entangled particle would be equivalent to using a
non-empty box in an inertial entanglement experiment.
Furthermore, the evolution depends on the way in which the cavity is 
accelerated -- e.g., whether it is subject to a continuously varying 
Lorentz length contraction (in the frame of the box) or not.
(The temperature experienced by Rob is only relevant if the cavity is  
not small compared to the characteristic length scale of Rob's 
trajectory.) 
If Rob's cavity is stationary with respect to  Rob's proper 
(i.e., Rindler) time, then the corresponding Fulling-Rindler vacuum
state inside the cavity is stable under time-evolution and 
(in this sense) there is no particle creation at all.
Otherwise, if the cavity is not stationary with respect to the Rindler
time, then one also has to take into account particle creation due to
the dynamical Casimir effect (non-inertially moving wall/mirror effect)
for non-adiabatic changes.

{\bf II.} Inside a perfectly conducting ideal lossless cavity 
(as assumed by the authors), the time-evolution of the quantum state
is always unitary and hence a pure state remains a pure state. 
Consequently, in this situation, any fidelity loss can only be induced
by an inappropriate measurement (e.g., measuring the wrong kind of
particle, see the point above) or by an imperfect generation of the
entangled pair at the moment when Alice and Rob coincide
(see also the point below). 
If the pair of photons is perfectly entangled initially and the
cavity is assumed to be ideal then there is always a measurement 
procedure (i.e., a set of positive/negative frequency basis functions)
which gives fidelity one.

{\bf III.} Formul{\ae} (6) and (7) in Ref.~\cite{Alsing} describe the
relation of the Minkowski and the Fulling-Rindler modes in a
space-time without any boundaries. 
Therefore, these expressions and hence also the subsequent equations  
do not describe the cavity modes in general.
The application of the thermo-field formalism to the case with
cavities requires more detailed considerations -- e.g., how 
does the quantum state within the cavity in the right Rindler wedge
$I$ get entangled with the state of the field in the left Rindler
wedge $II$.  
In general one can always set up the cavities so there is no
entanglement between the cavities in the two wedges. 
One could of course also set them up so as to have entanglement, by,
for example, opening each of the cavities to the Minkowski vacuum
outside the cavities for a while.   

In summary, the impact of the thermal nature (as experienced by Rob)
of the Minkowski vacuum on the fidelity would be expected to be an
issue  only if Rob attempts to make measurements on unconstrained
entangled photons (i.e., without cavities) emitted by Alice, 
for example.
E.g., if the frequency of the photons is not large compared to the 
temperature experienced by Rob, i.e., their wave-length is not much 
smaller than the characteristic length scale of Rob's trajectory, 
the photons cannot strictly be localized inside of Rob's horizon and, 
consequently, Rob cannot recover the full information sent by Alice.
In the case with cavities considered in Ref.~\cite{Alsing}, however, 
the results obtained there cannot be applied in full generality 
without additional considerations and a potential loss of
fidelity depends upon the various details discussed above. 

{\em Acknowledgments}
R.~S.~gratefully acknowledges fruitful discussions with U.~Fischer and 
financial support by the Emmy-Noether Programme of the German Research 
Foundation (DFG) under grant No.~SCHU 1557/1-1 
and by the Humboldt foundation.
Furthermore, this work was supported by the 
COSLAB Programme of the ESF, the CIAR, and the NSERC.
\\
\\
Ralf Sch\"utzhold$^1$ and William G.~Unruh$^2$

$^1$Institut f\"ur Theoretische Physik

Technische Universit\"at Dresden

01062 Dresden, Germany

$^2$Department of Physics and Astronomy

University of British Columbia

Vancouver B.C., V6T 1Z1 Canada
\\

\end{document}